\documentclass{appolb}
\usepackage{graphicx}
\usepackage{subfigure}

\preprint{}

\begin{document}

\title{{\bf Consistent hydrodynamic description of one- and two-particle observables in relativistic heavy-ion collisions at RHIC}~\thanks{Talk presented  by WF at the {\em IV Workshop on Particle Correlations and Femtoscopy}, Krak\'ow, Poland, Sept. 11 -- 14, 2008}~\thanks{Partly supported by the Polish Ministry of Science and Higher Education, grants N202 153 32/4247 and N202 034 32/0918, and by the U.S. National Science Foundation under Grant No. PHY-0653432.}}
\author{Wojciech Florkowski $^{1,2}$, Wojciech Broniowski $^{1,2}$, \\  Mikolaj Chojnacki $^1$ and Adam Kisiel $^{3,4}$ 
\address{
$^1$ H. Niewodnicza\'nski Institute of Nuclear Physics, Polish Academy of Sciences, PL-31342 Krak\'ow, Poland \\
$^2$ Institute of Physics, Jan Kochanowski University, PL-25406~Kielce,~Poland \\
$^3$ Faculty of Physics, Warsaw University of Technology, PL-00661 Warsaw, Poland \\
$^4$ Department of Physics, Ohio State University, 1040 Physics Research Building, 191 West Woodruff Ave., Columbus, OH 43210, USA}}

\maketitle

\begin{abstract}
We show that a consistent hydrodynamic description of soft-hadronic one- and two-particle observables (the HBT radii) studied in the relativistic heavy-ion collisions at RHIC may be obtained if one uses the Gaussian energy density profile as the initial condition. The transverse-momentum spectra, the elliptic flow coefficient $v_2$, and the pionic azimuthally sensitive HBT radii are successfully reproduced, which hints that the long standing HBT puzzle has been solved. 
\end{abstract}

 \medskip

The intermediate stages of relativistic heavy-ion collisions are commonly described in the framework of the hydrodynamic models \cite{Teaney:2000cw,Hirano:2002ds,Kolb:2003dz,Huovinen:2003fa,Eskola:2005ue,Hama:2005dz,Nonaka:2006yn}. Such models turned out to be very successful in reproducing the one-particle measurements such as the transverse-momentum spectra and the elliptic flow coefficient $v_2$. In this context, the early starting time of the hydrodynamic evolution that was needed to describe the data as well as the explanation of the quite large experimental values of $v_2$ were interpreted as the evidence that the quark-gluon plasma created at RHIC is a strongly interacting system~\cite{Shuryak:2004kh}. On the other hand, the hydrodynamic models failed in reproducing the experimental two-particle measurements, such as the pion correlation functions. The most distinct example is the ratio of the so called HBT radii $R_{\rm out}$ and $R_{\rm side}$ -- in the typical hydrodynamic calculations it comes out too large, exceeding the experimentally measured values by about 20-50\%.  

\begin{figure}[t]
\begin{center}
\includegraphics[angle=0,width=0.45 \textwidth]{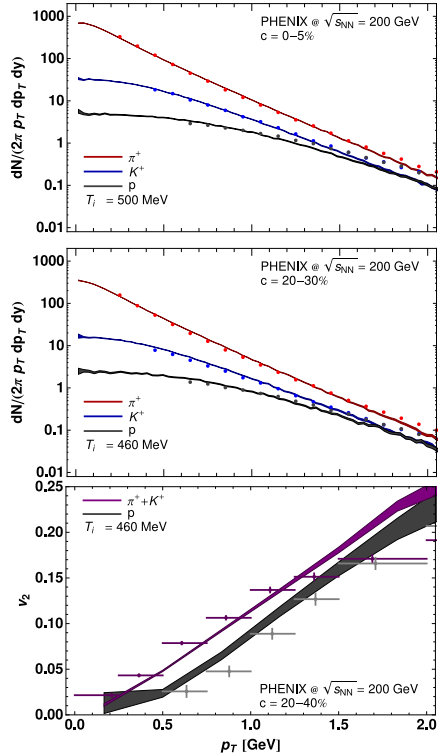}
\end{center}
%\vspace{-6.5mm}
\caption{The transverse-momentum spectra of pions, kaons and protons for the centrality bin $c$ = 0-5\% (upper panel), $c$ = 20-30\% (middle panel), and the elliptic flow coefficient $v_2$ for  $c$ = 20-40\% (lower panel), plotted as functions of the transverse momentum and compared to the RHIC Au+Au data \cite{Adler:2003cb,Adler:2003kt}.}
\label{fig:spv2}
\end{figure}

In our recent work \cite{Broniowski:2008vp}, we have found  that the consistent description of one- and two-particle observables in the soft region is achieved within the hydrodynamic model if one modifies the initial conditions -- the initial energy profile obtained in most cases from the optical Glauber model should be replaced by the Gaussian profile. Other ingredients of our model, such as the use of the semi-hard equation of state, are also important for the overall good agreement, however, taken into account alone they are not sufficient to get the satisfactory results. In this aspect, our findings are complementary to the observations made recently by Pratt (Ref. \cite{Pratt:2008qv} and these proceedings \cite{Pratt:2008bc}).

Our approach is based on the  2+1 boost-invariant inviscid hydrodynamics  \cite{Chojnacki:2004ec,Chojnacki:2006tv,Chojnacki:2007jc} followed by the statistical-hadronization model THERMINATOR \cite{Kisiel:2005hn}. The initial energy-density profiles are obtained from the Monte-Carlo Glauber model GLISSANDO \cite{Broniowski:2007nz}, which includes the eccentricity fluctuations \cite{Miller:2003kd,Bhalerao:2005mm,Voloshin:2006gz,Andrade:2006yh,Alver:2006wh}. The results of those calculations are approximated by the Gaussians. The thermal event generator THERMINATOR simulates hadron emission from the freeze-out hypersurface determined by the hydrodynamic calculation. The single freeze-out scenario \cite{Broniowski:2001we,Broniowski:2002nf} is assumed with the universal final temperature $T_f = $145 MeV\footnote{The relatively high decoupling temperature may be connected with the presence of the instabilities at the phase transition connected with the diverging bulk viscosity \cite{Torrieri:2008ip}.}. Besides the final temperature our model has essentially two additional parameters: the initial temperature $T_i$, fixing the absolute normalization, and the initial time for the start of hydrodynamics, \mbox{$\tau_0$ = 0.25 fm}. For each centrality class, the widths of the two-dimensional initial Gaussian energy distribution  are obtained directly from the GLISSANDO simulations. 

\begin{figure}[tb]
\begin{center}
\includegraphics[angle=0,width=0.95 \textwidth]{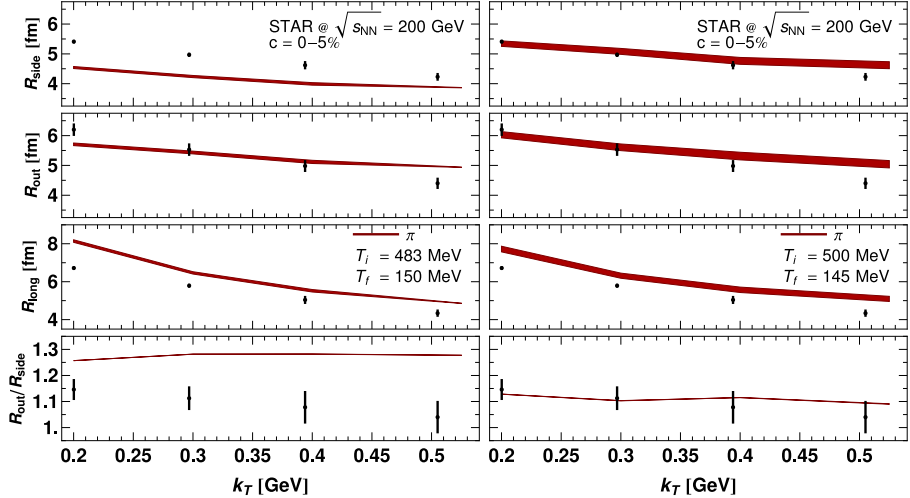}
\end{center}
%\vspace{-6.5mm}
\caption{ The pion HBT radii $R_{\rm side}$ , $R_{\rm out}$ , $R_{\rm long}$,  and the ratio $R_{\rm out}/R_{\rm side}$ for central collisions, shown as the functions of the average momentum of the pair and compared to the RHIC Au+Au data \cite{Adams:2004yc}. Left: our best results obtained with the Glauber initial condition \cite{Chojnacki:2007rq}. Right: our results obtained with the Gaussian initial conditions (i.e., with the Gaussian approximation to the the Glauber simulations) \cite{Broniowski:2008vp}.
\label{fig:hbt}}
\end{figure}

Figure \ref{fig:spv2} shows our results for the transverse-momentum spectra of pions, kaons, and protons for the centrality classes $c$ = 0-5\% and \mbox{$c$ = 20-30\%}. Fig.~\ref{fig:spv2} presents also our results for $v_2$ for the centrality \mbox{$c$ = 20-40\%}, plotted as functions of the transverse momentum. The spectra and the elliptic flow are compared to the RHIC data \cite{Adler:2003cb,Adler:2003kt}. We observe a very good agreement between the model predictions and the data. The small excess of the theoretical proton $v_2$ above the data may be attributed to the lack of rescattering in the final state. 

\begin{figure*}[tb]
\begin{center}
\includegraphics[angle=0,width=0.95 \textwidth]{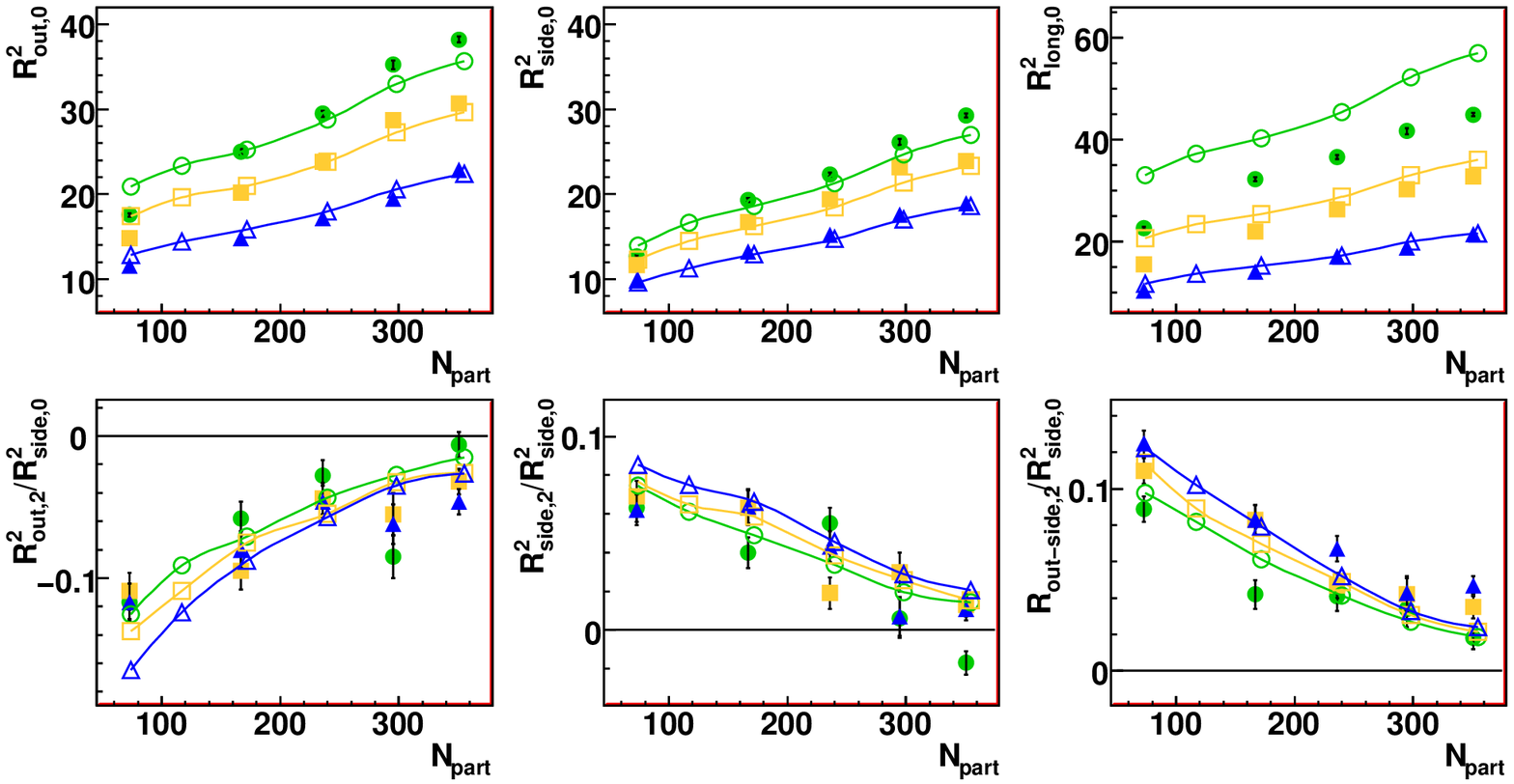}
\end{center}
\vspace{-6.5mm}
\caption{Results for the RHIC HBT radii and their azimuthal oscillations. For
each value of $N_{\rm part}$ on the horizontal axis the
experimental points (filled symbols) and the model results (empty
symbols) are plotted. The points from top to bottom at each plot correspond to
$k_T$ contained in the bins 0.15-0.25~GeV (circles),
0.25-0.35~GeV  (squares), and 0.35-0.6~GeV(triangles). The top panels
show $R^{2}_{\rm out,0}$, $R^{2}_{\rm side,0}$, and $R^{2}_{\rm
long,0}$, the bottom panels show the magnitude of the allowed
oscillations divided conventionally by $R_{\rm side,0}^2$. }
\label{fig:resultsrhic}
\end{figure*}

Figure \ref{fig:hbt} presents our results for the pion HBT radii $R_{\rm side}$ , $R_{\rm out}$, $R_{\rm long}$,  and the ratio $R_{\rm out}/R_{\rm side}$ for central collisions, compared to the RHIC data \cite{Adams:2004yc}. The left panel shows our best results obtained with the traditional Glauber initial condition \cite{Chojnacki:2007rq}, while the right panel shows the results obtained with the Gaussian initial condition \cite{Broniowski:2008vp}. One observes that a very good agreement between the data and the theoretical model predictions is achieved in the case where the Gaussian initial condition is used. We note that the calculation of the radii does not introduce any extra parameters. All the characteristics of the emitting source were already fixed by the fits to the spectra and $v_2$. Finally, in Fig. \ref{fig:resultsrhic} the results describing the azimuthal dependence of the HBT radii are plotted \cite{Kisiel:2008ws}. Here $R^2(\phi) = R_0^2 + 2 R_2^2 \cos(2\phi)$. Again, we observe a very good agreement between the data and our model for different centralities and different average momenta of the pion pairs $k_T$. 
 
Our results indicate that it is possible to achieve a consistent hydrodynamic description of one- and two-particle soft hadronic data collected in the RHIC heavy-ion experiments at the highest beam energies. In particular, it is possible to describe simultaneously the transverse-momentum spectra, the elliptic flow coefficient $v_2$, and the HBT correlation radii (the preliminary results show that the correlations of the non-identical particles are also well reproduced in our model). Our finding suggests that there exists a solution to the long standing RHIC HBT puzzle understood as the failure of the relativistic hydrodynamics  in reproducing consistently the spectra and the HBT radii. 

The Gaussian initial profile leads to a faster development of the initial transverse flow which makes the system evolution shorter. At the same time the transverse size of the system at freeze-out is slightly larger (as compared to the standard Glauber scenario). These two effects put together lead to the desired reduction of the ratio $R_{\rm out}/R_{\rm side}$. In the approach discussed in Refs. \cite{Pratt:2008qv,Pratt:2008bc}, the crucial role of the faster initial acceleration \cite{Chojnacki:2004ec,Gyulassy:2007zz} is also emphasized. However, in \cite{Pratt:2008bc} this effect comes from the very early start of the hydrodynamics, $\tau_0 = 0.1$, fm and from the inclusion of viscous terms which increase the transverse acceleration with respect to the longitudinal one. We stress, however, that our approach remains the only one which describes correctly the HBT radii and the elliptic flow, as the studies of Refs. \cite{Pratt:2008qv,Pratt:2008bc} are restricted to the cylindrically symmetric case.

Clearly, the issue of the relative importance of the {{\it initial energy profile}, the {\it initial pre-equilibrium flow}, and the {\it viscous effects} should be addressed in further investigations. The recent calculation \cite{Pratt:2008sz} shows that the pre-equilibrium flow is inevitable as it comes from the energy-momentum conservation laws. On the other hand, the specific amount of the pre-equilibrium flow required to correctly describe the data depends on the initial energy-density profile. In  \cite{Broniowski:2008qk} we have assumed that the start of the hydrodynamic evolution may be delayed to $\tau =$ 1 fm. The earlier evolution consists of the free-streaming (in the proper time interval: \mbox{0.25 fm $\leq \tau \leq$ 1 fm}) followed by the sudden equilibration and transition to the hydrodynamic regime (at $\tau =$ 1 fm). We have used again the Gaussian initial conditions and shown that they lead the sufficient amount of the pre-equilibrium transverse flow at $\tau = 1$ fm, when the hydrodynamic evolution starts, i.e., the FS+SE approach of Ref. \cite{Broniowski:2008qk} describes the data equally well as our standard approach. This behavior suggest some interesting universality of the Gaussian initial conditions. Of course, the open question remains to find the microscopic mechanism leading to such a form of the initial conditions.

\end{document}